\documentclass[conference,letterpaper]{IEEEtran}

\usepackage[utf8]{inputenc} 
\usepackage[T1]{fontenc}
\usepackage[font=footnotesize,labelfont=bf]{caption}
\usepackage{url}
\usepackage{ifthen}
\usepackage{cite}
\usepackage[cmex10]{amsmath} 
\usepackage[dvipsnames]{xcolor}
\usepackage{changepage}
\usepackage{tikz}
\usetikzlibrary{calc}
\usepackage{color,hyperref,listings}
\usepackage[shortlabels]{enumitem}
\usepackage{algorithm}
\usepackage{algorithmic}
\usepackage{amssymb}
\usepackage{tkz-euclide}
\usepackage{siunitx}
\usepackage{physics}
\usepackage{multicol}
\usepackage{pgfplots}
\usepackage{adjustbox}
\usepackage{cancel}
\usepackage{commath}
\usepackage{graphicx}
\usepackage{mathdots}
\usepackage{tabularx}
\usepackage{mathtools}
\usepackage{booktabs}
\usepackage{environ}
\usepackage{cleveref}
\usepackage{fancyhdr}
\usepackage{centernot}
\usepackage[margin=0.75in]{geometry}
\usepackage[american,siunitx,arrowmos]{circuitikz}
\usepackage[final]{pdfpages}
\pgfplotsset{compat=1.18}
\usepackage[english]{babel}
\usepackage{fontawesome}
\usepackage{listings}
\usepackage{fixmath}
\usepackage{xspace}

\usepackage{listings}
\usepackage{amsthm}
\usepackage{upgreek}
\usetikzlibrary{trees}
\usepackage{amsmath,amsfonts}
\usepackage{array}
\usepackage{subcaption}
\usepackage{textcomp}
\usepackage{stfloats}
\usepackage{url}
\usepackage{verbatim}
\usepackage{cite}
\usepackage{graphicx}
\usepackage{prettyref}
\usepackage{wrapfig}
\newcommand{\Ebb}{\mathbb{E}}

\newcommand{\yv}{\mathbf{y}}

\newcommand{\Acal}{\mathcal{A}}

\newcommand{\Ccal}{\mathcal{C}}

\newcommand{\Scal}{\mathcal{S}}

\renewcommand{\Gamma}{\Upgamma}
\renewcommand{\Theta}{\Uptheta}
\renewcommand{\Omega}{\Upomega}

\newcommand{\convas}{\stackrel{\text{a.s.}}{\longrightarrow}}

\newcommand{\indic}[1]{\mathbf{1}\left\{#1\right\}}

\newcommand{\simiid}{\stackrel{\text{i.i.d.}}{\sim}}

\def\ie{\textit{i.e.}\@\xspace}

\newtheorem{theorem}{Theorem}[section]

\newtheorem{corollary}[theorem]{Corollary}
\newtheorem{fact}[theorem]{Fact}

\theoremstyle{definition}
\newtheorem{definition}[theorem]{Definition}

\newtheorem{exercise-easy}[theorem]{Exercise}
\newtheorem{exercise-med}[theorem]{Exercise}
\newtheorem{exercise-hard}[theorem]{Exercise$^\star$}

\newtheorem*{claim*}{Claim}

\newtheorem{remark}[theorem]{Remark}
\newtheorem*{remark*}{Remark}

\newtheorem*{observation*}{Observation}

\newcommand{\savehyperref}[2]{\texorpdfstring{\hyperref[#1]{#2}}{#2}}

\newrefformat{eq}{\savehyperref{#1}{\textup{(\ref*{#1})}}}
\newrefformat{ineq}{\savehyperref{#1}{\textup{(\ref*{#1})}}}
\newrefformat{eqn}{\savehyperref{#1}{\textup{(\ref*{#1})}}}
\newrefformat{lem}{\savehyperref{#1}{Lemma~\ref*{#1}}}
\newrefformat{con}{\savehyperref{#1}{Construction~\ref*{#1}}}
\newrefformat{def}{\savehyperref{#1}{Definition~\ref*{#1}}}
\newrefformat{thm}{\savehyperref{#1}{Theorem~\ref*{#1}}}
\newrefformat{cor}{\savehyperref{#1}{Corollary~\ref*{#1}}}
\newrefformat{cha}{\savehyperref{#1}{Chapter~\ref*{#1}}}
\newrefformat{sec}{\savehyperref{#1}{Section~\ref*{#1}}}
\newrefformat{app}{\savehyperref{#1}{Appendix~\ref*{#1}}}
\newrefformat{tab}{\savehyperref{#1}{Table~\ref*{#1}}}
\newrefformat{fig}{\savehyperref{#1}{Figure~\ref*{#1}}}
\newrefformat{hyp}{\savehyperref{#1}{Hypothesis~\ref*{#1}}}
\newrefformat{alg}{\savehyperref{#1}{Algorithm~\ref*{#1}}}
\newrefformat{rem}{\savehyperref{#1}{Remark~\ref*{#1}}}
\newrefformat{item}{\savehyperref{#1}{Item~\ref*{#1}}}
\newrefformat{step}{\savehyperref{#1}{step~\ref*{#1}}}
\newrefformat{conj}{\savehyperref{#1}{Conjecture~\ref*{#1}}}
\newrefformat{fact}{\savehyperref{#1}{Fact~\ref*{#1}}}
\newrefformat{prop}{\savehyperref{#1}{Proposition~\ref*{#1}}}
\newrefformat{prob}{\savehyperref{#1}{Problem~\ref*{#1}}}
\newrefformat{claim}{\savehyperref{#1}{Claim~\ref*{#1}}}
\newrefformat{relax}{\savehyperref{#1}{Relaxation~\ref*{#1}}}
\newrefformat{rem}{\savehyperref{#1}{Remark~\ref*{#1}}}
\newrefformat{red}{\savehyperref{#1}{Reduction~\ref*{#1}}}
\newrefformat{part}{\savehyperref{#1}{Part~\ref*{#1}}}
\newrefformat{ex}{\savehyperref{#1}{Exercise~\ref*{#1}}}
\newrefformat{property}{\savehyperref{#1}{Property~\ref*{#1}}}
\newrefformat{type}{\savehyperref{#1}{Type~\ref*{#1}}}
\newrefformat{eg}{\savehyperref{#1}{Example~\ref*{#1}}}
\newrefformat{obs}{\savehyperref{#1}{Observation~\ref*{#1}}}

\definecolor{deepblue}{rgb}{0,0,0.5}
\definecolor{deepred}{rgb}{0.6,0,0}
\definecolor{deepgreen}{rgb}{0,0.5,0}
\begin{document}
\title{LZMidi: Compression-Based Symbolic Music Generation} 

\author{%
  \IEEEauthorblockN{Connor Ding, Abhiram Gorle, Sagnik Bhattacharya, Divija Hasteer, Naomi Sagan, Tsachy Weissman}
  \IEEEauthorblockA{Department of Electrical Engineering, Stanford University\\
                    Stanford, CA, 94305 \\
                    czsding, abhiramg, sagnikb, dhasteer, nasagan, tsachy@stanford.edu}

}


\maketitle

\begin{abstract}
   Recent advances in symbolic music generation primarily rely on deep learning models such as Transformers, GANs, and diffusion models. While these approaches achieve high-quality results, they require substantial computational resources, limiting their scalability. We introduce LZMidi, a lightweight symbolic music generation framework based on a Lempel-Ziv (LZ78)-induced sequential probability assignment (SPA). By leveraging the discrete and sequential structure of MIDI data, our approach enables efficient music generation on standard CPUs with minimal training and inference costs. Theoretically, we establish universal convergence guarantees for our approach, underscoring its reliability and robustness. Compared to state-of-the-art diffusion models, LZMidi achieves competitive Fréchet Audio Distance (FAD), Wasserstein Distance (WD), and Kullback-Leibler (KL) scores, while significantly reducing computational overhead—up to 30× faster training and 300× faster generation. Our results position LZMidi as a significant advancement in compression-based learning, highlighting how universal compression techniques can efficiently model and generate structured sequential data, such as symbolic music, with practical scalability and theoretical rigor.


\end{abstract}

\vspace{-10pt}

\section{Introduction}

\vspace{-10pt}

Deep learning–based generative models have achieved remarkable success in text, image, and audio synthesis. However, their substantial computational demands—particularly in sampling procedures such as those employed in diffusion-based models—pose significant challenges for practical deployment because of high latency and the reliance on specialized hardware. Recent research has explored more computationally tractable alternatives that maintain competitive output quality. \cite{sagan2024familylz78baseduniversalsequential} introduces a learning framework based on universal sequential probability assignments (SPAs) derived from the celebrated Lempel-Ziv (LZ78) \cite{1055934} compression. Using LZ parsing under \textit{stationary} and \textit{ergodic} assumptions, the proposed approach represents sequences efficiently within a tree-based structure. This methodology offers strong theoretical guarantees on runtime, memory usage and has demonstrated promising results in text generation, achieving low-latency training and sampling. 

Our present work focuses on the generation of symbolic music, which refers to the task of generating music in a structured, discrete format, typically represented as MIDI or other symbolic encodings rather than raw audio waveforms. With its discrete structure and finite alphabet, symbolic music is well-suited to LZ-based SPAs. In this work, we induce an LZ78-based SPA on symbolic music from the Lakh MIDI dataset and use this as a tool for symbolic music generation. Empirical evaluations indicate that LZ78-based SPA produces music of excellent perceptual quality—quantified using Fréchet Audio Distance (FAD)—while significantly reducing both training time and sampling overhead.  

Notably, our proposed framework operates efficiently on standard CPUs, eliminating the need for energy-intensive GPUs. This not only enhances accessibility for researchers and practitioners with limited computational resources but also aligns with sustainability goals by reducing environmental impact. Collectively, these results position LZMidi as a compelling, resource-efficient alternative to deep learning approaches in symbolic music generation.

\vspace{-10pt}

\section{Related Work}

\vspace{-10pt}

\subsection{Deep Learning for MIDI Generation} 

\vspace{-10pt}

State-of-the-art symbolic music generation predominantly employs deep learning techniques, with transformer-based models and diffusion frameworks leading recent advancements. Early influential contributions within the \href{https://magenta.tensorflow.org/research}{Magenta} suite, such as MusicVAE \cite{roberts2019hierarchicallatentvectormodel}, provided hierarchical VAEs for capturing long-term musical structure, while Music Transformer \cite{huang2018musictransformer} introduced relative attention mechanisms to handle long-range dependencies. Transformer-GANs \cite{Muhamed_Li_Shi_Yaddanapudi_Chi_Jackson_Suresh_Lipton_Smola_2021} integrate Transformers with GAN architectures to enhance sequence coherence and mitigate exposure bias. 

Concurrently, diffusion-based works, such as \cite{mittal2021symbolicmusicgenerationdiffusion} and \cite{plasser2023discretediffusionprobabilisticmodels}, advanced symbolic music generation through iterative refinement procedures in continuous and discrete latent spaces, respectively. Although effective, these approaches remain computationally intensive. Recent efforts, including fast diffusion GANs \cite{zhang2024composerstylespecificsymbolicmusic}, attempt to accelerate generation by reducing denoising steps. Motivated by these computational constraints, our work introduces compression-based universal sequential probability assignments (SPA) as a lightweight, scalable alternative achieving comparable generative quality with significantly reduced training and sampling costs.

\vspace{-10pt}

\subsection{Universal Information Processing for MIDI data} 

\vspace{-10pt}


Prior to the deep learning era, universal algorithms such as Lempel-Ziv (LZ77, LZ78) \cite{1055934, 1055714} and Context-Tree Weighting (CTW) \cite{382012} inspired extensive research in symbolic music processing. This includes (1) identifying themes and patterns in musical data, (2) compression-based similarity and perceptual measures, and (3) sequence-prediction algorithms for music generation \cite{Pearce03042017, conklin1995multiple, rolland2001modeling}. In particular, in the context of symbolic music generation, \cite{begleiter2004prediction} employs CTW to model musical event probabilities based on prior context, effectively capturing hierarchical and temporal dependencies. This approach directly inspires our use of LZ78-based sequential probability assignment. Additionally, \cite{langdon1983note} also hints at the possible efficacy of LZ78 for universal sequence modeling, which further motivates our efforts. 

\vspace{-10pt}

\subsection{Compression for Learning}

\vspace{-10pt}

Compression serves as a proxy for learning because an algorithm’s ability to compress data reflects its capacity to capture underlying structure. \cite{dele2024language} argues that language modeling is equivalent to compression, as minimizing the expected negative log-likelihood minimizes the expected code length. In essence, a model with lower perplexity compresses text more efficiently, demonstrating its grasp of statistical regularities. Moreover, off‐the‐shelf compressors can function as probabilistic sequence models by selecting the token that minimizes compressed file size. Similarly, compressor-based methods, which operate without explicit model parameters, have demonstrated competitive performance in classification tasks, highlighting that compression itself can leverage inherent redundancies to approximate learned representations without conventional model training \cite{jiang2023lowresource}. \footnote{However, we note that the validity of such compressor-based classification methods has sparked some debate regarding their empirical accuracy evaluation criteria and whether such accuracy comparisons might be inherently biased or overly optimistic.} Together, these results imply that minimizing redundancy is tantamount to learning the data's probabilistic structure

Further, Merhav and Weinberger \cite{1262613} examine "universal simulation", a closely related concept demonstrating that sampling uniformly from the type class of a training sequence yields samples exactly matching the source distribution, thus minimizing dependency (mutual information) with the training data. They quantify this dependency as the "price of universality", reflecting the statistical cost incurred when generating samples without knowledge of the source distribution. Our LZMidi method is directly motivated by these foundational concepts: it leverages universal compression (via LZ78-based SPA) to efficiently approximate the underlying statistical structure of symbolic music. This allows LZMidi to effectively capture the intrinsic redundancy and repetitive structure in musical sequences \cite{Chen2023QuantifyingRI}, providing a resource-efficient alternative for symbolic music generation.



\vspace{-10pt}

\section{Theoretical Background}

\vspace{-10pt}




In this section, we introduce the LZ78-based sequential probability assignment (SPA) from \cite{sagan2024familylz78baseduniversalsequential} and outline a theoretical justification of its application to symbolic music. Consider a finite, individual sequence $x^n = (x_1, x_2, \dots, x_n)$, where each symbol $x_i$ takes a value from a finite alphabet $\mathcal{X}$.
The tree-based Lempel-Ziv algorithm (LZ78) \cite{1055934} is a universal compression algorithm where the number of bits expended per source symbol converge to the entropy rate, while also
inducing a sequential probability assignment. Following this idea, \cite{sagan2024familylz78baseduniversalsequential} derives a practical sequential probability assignment: for a given sequence $x^{t-1}=(x_1, x_2,\dots, x_{t-1})$, to predict the next symbol $x_t$, we can derive the probability of observing symbol $a$ given context $x^{t-1}$ shown below.
We adopt the notations from \cite{sagan2024familylz78baseduniversalsequential}:
\begin{equation}
    q^{LZ,\gamma}(a|x^{t-1})\triangleq \frac{N_{LZ}(a|x^{t-1})+\gamma}{\sum_{a'\in \mathcal{X}}N_{LZ}(a'|x^{t-1})+\gamma|\mathcal{X}|}.
    \label{eq:spa}    
\end{equation}
The probability assignment for symbol $a$ given context $x^{t-1}$ identifies the fraction of times symbol $a$ followed the context as $x_t$; it includes a perturbation term directed by $\gamma$ to tune how much the probability assignment should respect the empirical distribution from training data. Further details regarding the outline of the LZ78-induced SPA are provided in Appendix \ref{app:SPA}.
To justify our use of the LZ tree model, we present the following theorem, which establishes that an LZ tree trained on a sufficiently large dataset will closely approximate the true data distribution: 
\begin{theorem}[Universal Convergence of LZ78-SPA]
\label{thm:LZ78-main}
Let \(P\) be the law of a process with components taking values in a finite alphabet \(\mathcal{X}\), and let \(Q^m\) be the LZ78-based sequential probability assignment (SPA) constructed using \(m\) i.i.d training sequences from \(P_{X^n}\). Then, for any fixed \(n\),
\[
D\bigl(P_{X^n} \,\big\|\, Q^m_{X^n}\bigr) \;\xrightarrow[m\to\infty]{\text{a.s.}}\; 0,
\]
where \(D(\cdot\|\cdot)\) denotes the Kullback--Leibler divergence.
\end{theorem}

\emph{Proof Sketch.} By construction, each node in the LZ78 tree tracks the empirical frequency of symbols following a particular context, and since every context with nonzero probability is visited infinitely often, the frequency with which a symbol \(a\) appears converges to the true conditional probability \(P(a | \text{context})\). Consequently, the LZ78-SPA, which returns $q(a|\textrm{context})$, an estimate of $P$ via empirical frequencies, assigns probabilities that  approximate $P$ increasingly accurately with more training data. As a result, when the assigned probabilities agree with the source distribution for all positively-probable contexts, the relative entropy \(D(P_{X^n}\|Q^m_{X^n})\) converges to 0 almost surely as \(m\to\infty\).


A full, detailed proof is provided in Appendix~\ref{app:proof}. Here, we emphasize that the key idea relies on the law of large numbers, which can be invoked because each relevant context is visited infinitely often. This allows us to conclude that local (node-wise) empirical distributions converge to the true underlying source probabilities.

\begin{remark}
While standard universal compression theory typically relies on ergodicity for asymptotic guarantees, our practical setting employs fixed-length sequences, making explicit ergodicity assumptions less critical for our empirical results.
\end{remark}

\vspace{-10pt}

\section{Methods}

\vspace{-10pt}

\subsection{Data Structure}

\vspace{-10pt}

\begin{figure}[htbp]
    \centering
    \includegraphics[width=1\linewidth]{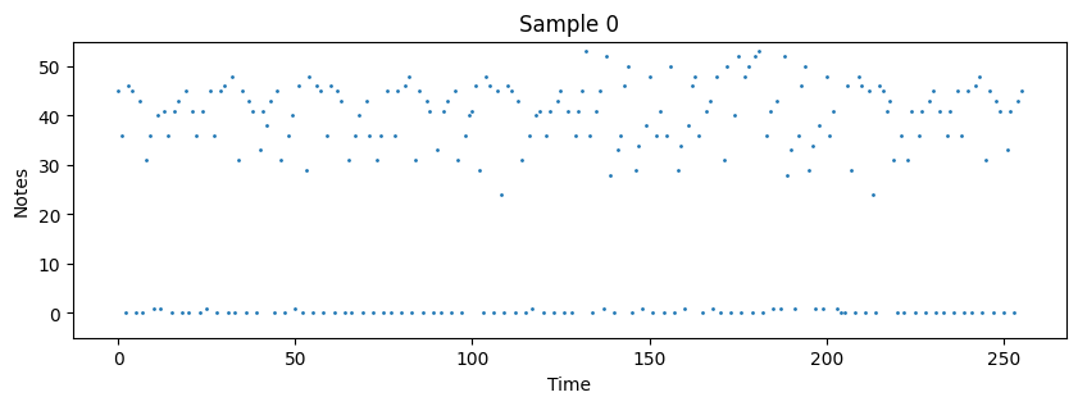}
    \caption{Sample Midi File}
    \label{fig:sample}
\end{figure}

\begin{figure}[htbp]
    \centering
    \begin{minipage}[b]{0.23\textwidth}
        \centering
        \includegraphics[width=\textwidth]{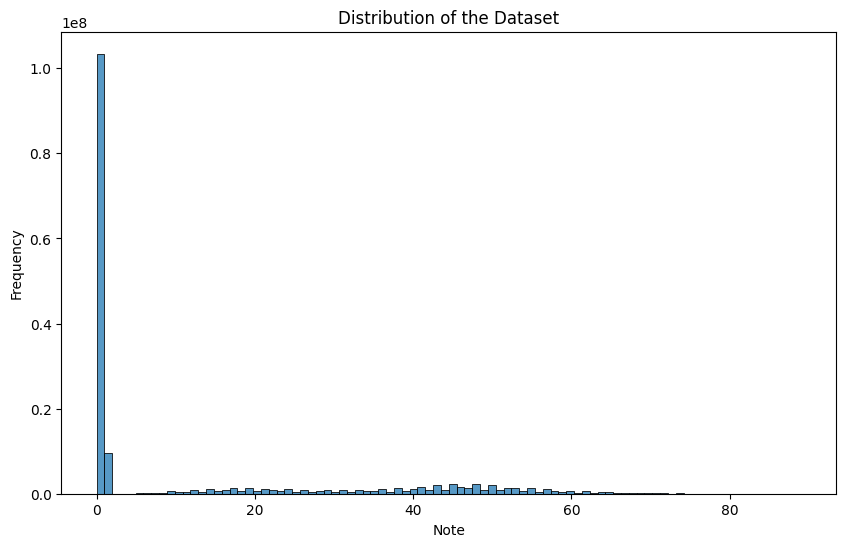}
        \caption{Data Distribution including 0 \& 1}
        \label{fig:data_dist}
    \end{minipage}
    \hfill
    \begin{minipage}[b]{0.23\textwidth}
        \centering
        \includegraphics[width=\textwidth]{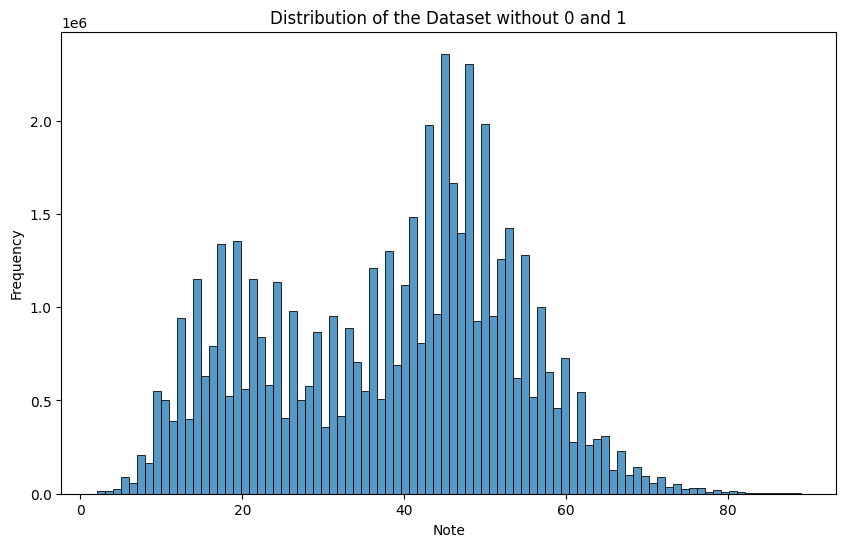}
        \caption{Data Distribution without 0 \& 1}
        \label{fig:data_dist_no01}
    \end{minipage}
\end{figure}

We use the \textbf{Lakh MIDI Dataset (LMD)}, containing 648,574 samples, each with 256 notes drawn from the alphabet \( \mathcal{X} = \{0, 1, \dots, 89\} \), to train our LZ-based model for symbolic music generation. Here, 0 represents a rest, 1 denotes consecutive note continuation, and 2–89 correspond to actual pitch values. Figure \ref{fig:sample} illustrates a sample MIDI sequence. The dataset's note distribution (Figure \ref{fig:data_dist}) highlights the dominance of rests (0s) and continuations (1s), while Figure \ref{fig:data_dist_no01} shows that actual note pitches are clustered around the mid-range. To build the alphabet for the LZ model, we simply treat each individual note as a symbol within the alphabet $\mathcal{X} = \{0, 1, 2, \dots, 89\}$, allowing us to traverse the tree and update the SPA sequentially for each note in the dataset. 



\subsection{Baseline Model}

\vspace{-10pt}

For comparison, we use the absorbing state discrete diffusion denoising model (\textbf{ASD3PM}) by Plasser et al. \cite{plasser2023discretediffusionprobabilisticmodels}, a discrete diffusion-based generative model for symbolic music. ASD3PM outperforms transformer-based autoregressive models \cite{choi2020encodingmusicalstyletransformer} and prior continuous diffusion approaches \cite{mittal2021symbolicmusicgenerationdiffusion}.
ASD3PM follows an iterative denoising process, progressively refining a noisy symbolic sequence. Unlike continuous diffusion models that rely on latent representations, ASD3PM directly models discrete token transitions, making it well-suited for symbolic data like MIDI. The forward process masks tokens using a transition matrix \( Q_t \), introducing an absorbing state for corrupted inputs, while the reverse process predicts \( p_\theta(x_0 | x_t) \) using a neural network. Inspired by \cite{austin2021structured}, it simplifies training by parameterizing loss directly on the original data \( x_0 \) and dynamically adjusting diffusion steps for flexible sampling. Its hierarchical architecture combining convolutional and transformer layers for refinement helps it outperform larger latent-space diffusion models \cite{mittal2021symbolicmusicgenerationdiffusion}.

The loss function minimizes the evidence lower bound (ELBO) on the likelihood of the original data:
\begin{equation}
\mathcal{L}_{\text{ELBO}} = \mathbb{E}_{q(x_0)}\left[\sum_{t=1}^T \frac{T - t - 1}{T} \mathbb{E}_{q(x_t|x_0)} \left[ \log p_\theta(x_0|x_t) \right]\right]
\end{equation}

where $T$ is the number of diffusion steps. This formulation ensures efficient training by focusing on intermediate diffusion steps. Its hierarchical architecture improves efficiency by reducing trainable parameters while maintaining fidelity and diversity.

Note that while ASD3PM is a state-of-the-art discrete diffusion model, it was originally designed for more elaborate tasks (e.g., polyphonic generation with flexible infilling). Our LZMidi focuses on simpler unconditional generation. 


\vspace{-10pt}

\subsection{Metrics}

\vspace{-10pt}

We will evaluate the quality of the generated music through the following set of metrics:

\subsubsection{Framewise Self-Similarity Metrics (Consistency and Variance)}

To evaluate statistical similarity between generated and original sequences, we adopt the overlapping area (OA) metric from \cite{mittal2021symbolicmusicgenerationdiffusion}, which quantifies local pitch and duration distributions. Using a sliding 4-measure window (with a 2-measure hop), we fit Gaussian PDFs to pitch (\( p(k) \)) and duration (\( d(k) \)) distributions. The overlapping area (OA) between adjacent windows is defined as:
\begin{equation}
\text{OA}(k, k+1) = 1 - \text{erf}\left( \frac{c - \mu_1}{\sqrt{2}\,\sigma_1} \right) + \text{erf}\left( \frac{c - \mu_2}{\sqrt{2}\,\sigma_2} \right)
\end{equation}
where \( c \) is the intersection of the two Gaussian PDFs, and \( (\mu_1,\sigma_1), (\mu_2,\sigma_2) \) are the respective means and standard deviations.

From the overlapping areas for pitch (\(\text{OA}_P\)) and duration (\(\text{OA}_D\)), we compute the \textit{consistency} (C) and \textit{variance} (Var) as follows:
\begin{equation}
\text{C} = \max\left(0, 1 - \frac{|\mu_\text{OA} - \mu_\text{GT}|}{\mu_\text{GT}}\right)
\end{equation}
\begin{equation}
\text{Var} = \max\left(0, 1 - \frac{|\sigma_\text{OA}^2 - \sigma_\text{GT}^2|}{\sigma_\text{GT}^2}\right)
\end{equation}
where \( \mu_\text{OA}, \sigma_\text{OA}^2 \) and \( \mu_\text{GT}, \sigma_\text{GT}^2 \) are the means and variances of generated and ground-truth samples, respectively.

\textit{Consistency} measures alignment with ground truth, while \textit{variance} reflects diversity. Higher consistency suggests realistic sequence structure, while balanced variance prevents mode collapse. However, strong OA scores alone \textbf{do not} guarantee perceptual quality, necessitating complementary evaluation (e.g., FAD). This motivates us to experiment with alternative qualitative metrics described below.

\vspace{-10pt}

\subsubsection{Fréchet Audio Distance (FAD)}
Fréchet Audio Distance (FAD), inspired by the Fréchet Inception Distance (FID) \cite{heusel2018ganstrainedtimescaleupdate}, quantifies how closely the statistical distribution of generated audio aligns with real data. It computes the Fréchet distance between feature embeddings extracted from a pre-trained model (e.g., VGGish \cite{7952261}). Lower FAD scores indicate better perceptual similarity, making it a robust metric for evaluating generative quality.

\vspace{-10pt}

\subsubsection{KL-Divergence}

Kullback-Leibler (KL) divergence measures the discrepancy between the probability distributions of real and generated data. While lower KL values suggest better alignment, it primarily favours distributional similarity over perceptual quality. Given that KL can favor models producing mode-collapsed outputs, we emphasize FAD as the primary metric for evaluating generation fidelity.

\vspace{-10pt}

\subsubsection{Wasserstein Distance}
While consistency, variance, FAD, and KL divergence target audio-based evaluations, we also require a metric that directly examines numerical sequence distributions—especially for hyperparameter tuning without relying on costly neural-network inferences. We therefore use the Wasserstein Distance (WD) \cite{arjovsky2017wasserstein, gulrajani2017improved}, which measures the minimal cost of transforming one distribution into another. In our setup, we compare feature distributions from real and generated data; with a lower WD indicating a closer match between generated outputs and the ground truth.


\vspace{-10pt}

\section{Experiments}

\vspace{-10pt}

\subsection{Training and Generation Setup}

\vspace{-10pt}

We trained the LZMidi model using the provided implementation of LZ-based SPA from \cite{sagan2024familylz78baseduniversalsequential}. The \textit{Lakh MIDI Dataset} was split into an \textbf{80/20} train-test split. For training, we iterated through all samples in the training set, updating the LZ tree with each sequence. For evaluation, we generated 1,000 samples for each value of block length. All experiments were conducted on a CPU (Apple M1 Chip, 2021 MacBook). 
Following are some key parameters we can fine-tune for our training and generation of the LZ: 
\vspace{-8pt}
\begin{enumerate}
    \item \textbf{Dirichlet Parameter} ($\gamma$): This parameter, used in the computation of the sequential probability assignment (Equation \ref{eq:spa}), determines the proximity of the SPA to the empirical distribution. A smaller $\gamma$ results in the SPA being closer to the empirical distribution. 
    \item \textbf{Top-$K$}: The number of allowed symbols from which the model predicts. 
    \item \textbf{Temperature}: This parameter controls the randomness of the generated output by adjusting the probabilities of predicted symbols. A value approaching $0$ samples the most likely outcome; a value approaching $1$ samples directly from the SPA; a value approaching $\infty$ samples from a uniform distribution over the symbols.
    \item \textbf{Minimum Context}: The minimum context length that the SPA maintains during prediction. We set this value to 64 in our experiment.
\end{enumerate}

\vspace{-8pt}

We performed a hyperparameter sweep using Optuna \cite{akiba2019optunanextgenerationhyperparameteroptimization} to find the $\gamma$, Top-$K$, and temperature that optimizes Wasserstein distance. We suggest a categorical selection of the hyperparameters to Optuna for the hyperparameter sweep with the Wasserstein distance as the objective to minimize. We observed that $\gamma$ and temperature affect the generation quality the most, with a smaller $\gamma\approx 5\times 10^{-5}$ and a larger temperature $T\approx 0.8$ being optimal in terms of the Wasserstein distance. For the final generation, we choose $\gamma = 5\times 10^{-5}, T = 0.8, $ and $K=8$ to be our selection of hyperparameters. 

    
    
To generate a sample, we randomly select a symbol from the alphabet as seed data, corresponding to a direct child node of the LZ root, and generate a sequence of length $256$. The generated sequence is then mapped from its integer representation to note values and post-processed to ensure that no `1' follows `0'. Finally, the sequence is plotted and converted into both MIDI and WAV formats for metric computation and audio analysis.

\begin{figure}[htbp]
    \centering
    \includegraphics[width=\linewidth]{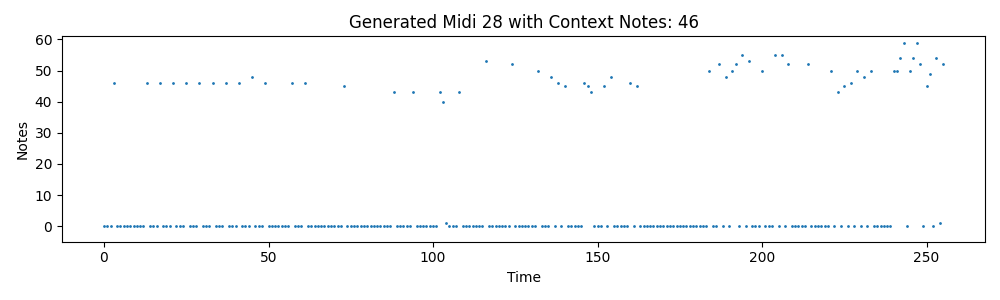}
    \vspace{2mm} 
    \includegraphics[width=\linewidth]{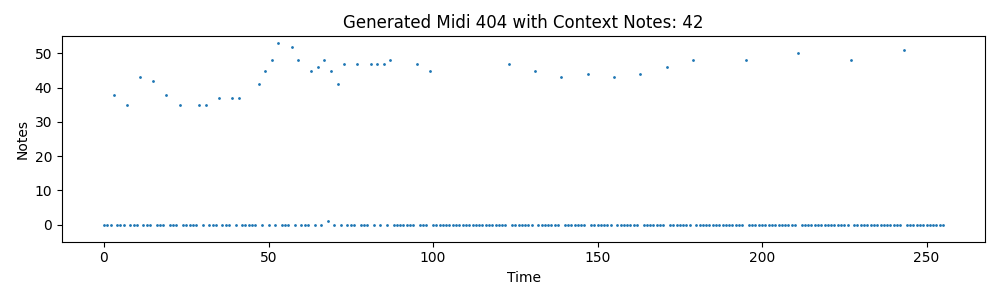}
    \vspace{2mm}
    \includegraphics[width=\linewidth]{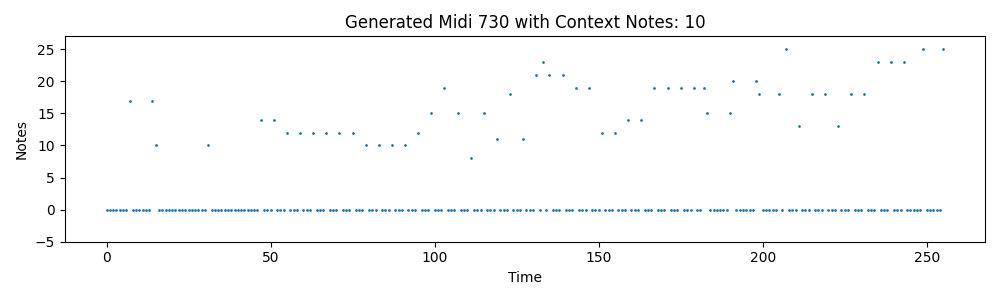}
    \caption{MIDI plots for Generated Samples using the LZMidi Model.}
    \label{fig:sample_all}
\end{figure}

Note that we independently trained the ASD3PM baseline due to significant differences in sequence length—our setup uses sequences of 256 tokens, whereas Plasser et al. \cite{plasser2023discretediffusionprobabilisticmodels} trained on 1024-token sequences—making direct use of these results unsuitable for a fair comparison. \cite{plasser2023discretediffusionprobabilisticmodels} reports a 24-hour training duration on 4x NVIDIA 2080 Ti GPUs. Due to computational constraints, we fixed our training time to approximately one hour. We include the metrics of the fully-trained model in the Appendix \ref{sec:fullytrained}.


\begin{figure}[htbp]
    \centering
    \includegraphics[width=\linewidth]{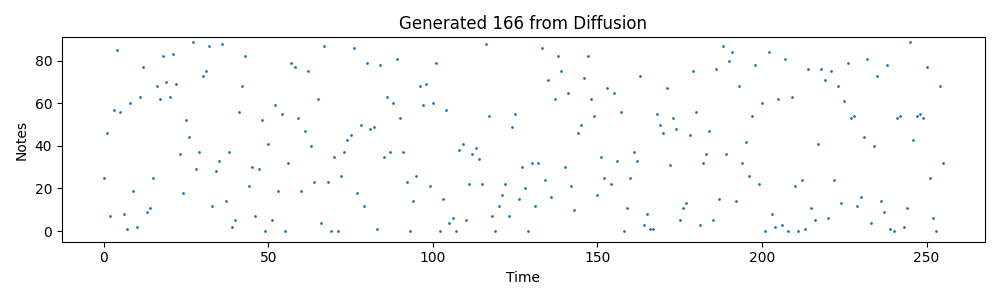}
    \vspace{2mm} 
    \includegraphics[width=\linewidth]{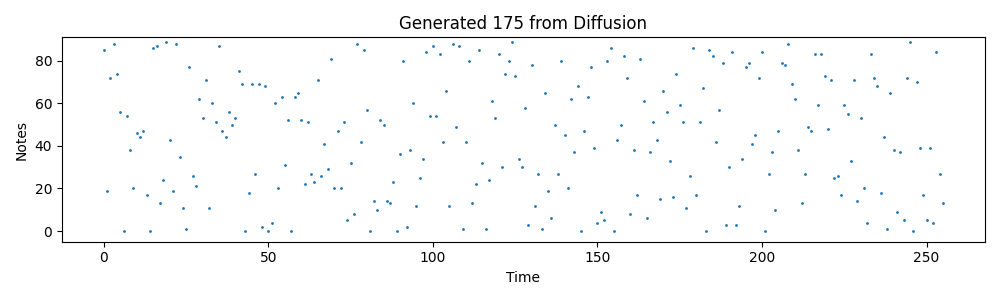}
    \vspace{2mm}
    \includegraphics[width=\linewidth]{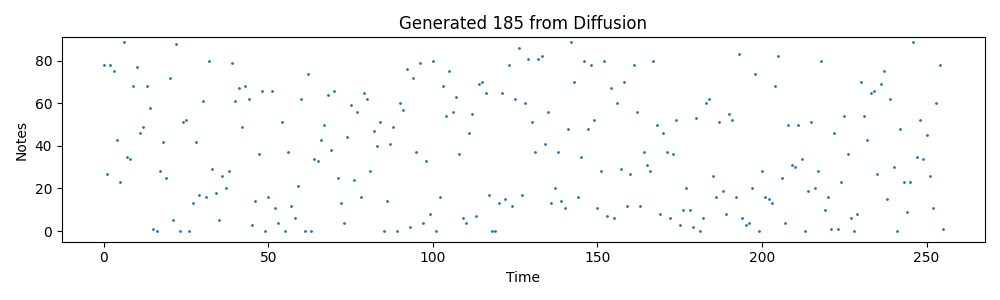}
    \caption{MIDI plots for Generated Samples using the D3PM Model.}
    \label{fig:diff_sample_all}
\end{figure}

\vspace{-8pt}

\subsection{Results}
\label{sec:results}

\vspace{-10pt}

We computed the consistency and variance metrics by comparing generated samples with 1000 randomly sampled sequences from both the training and test sets (in Table~\ref{tab:c_var_metrics}). LZMidi exhibits high consistency and variance for both pitch and duration, closely matching the dataset statistics. In fact, variance is greater in all cases for LZMidi.


\begin{table}[htbp]
\setlength{\tabcolsep}{5pt}
\caption{Consistency and Variance}
\label{tab:c_var_metrics}
    \begin{tabular}{|c|c|c|c|c|c|c|c|c|} \hline 
         &  \multicolumn{4}{|c|}{Training Set}& \multicolumn{4}{|c|}{Test Set}\\ \hline 
 & \multicolumn{2}{|c|}{Pitch}& \multicolumn{2}{|c|}{Duration}& \multicolumn{2}{|c|}{Pitch}& \multicolumn{2}{|c|}{Duration}\\ \hline 
 & C& Var&  C&Var&  C&Var&  C&Var\\\hline
         LZMidi&  $0.97$&  $\textbf{0.92}$&  $0.97$&$\textbf{0.93}$&  $0.97$&$\textbf{0.93}$& $0.97$&$\textbf{0.94}$\\ \hline 
 ASD3PM& $\textbf{0.98}$& $0.85$& $\textbf{0.99}$& $0.87$& $\textbf{0.98}$& $0.86$& $\textbf{0.99}$&$0.87$\\\hline
    \end{tabular}
\end{table}

We evaluate the quality of the generated MIDI samples using the three aforementioned metrics: Wasserstein Distance (WD), Fréchet Audio Distance (FAD), and Kullback-Leibler (KL) Divergence.  As shown in Table \ref{tab:fad_kl_metrics}, we compare our generated data to both the training and testing datasets. 
LZMidi achieves a much lower Wasserstein Distance (WD) and Fréchet Audio Distance (FAD) in both the training and test sets, indicating superior fidelity and distributional alignment.

\vspace{-10pt}

\begin{table}[htbp]
    \centering
\caption{WD, FAD and KL Divergence metrics}
\label{tab:fad_kl_metrics}
    \begin{tabular}{|c|c|c|c|c|c|c|}\hline
 & \multicolumn{3}{|c|}{Training}&  \multicolumn{3}{|c|}{Test}\\\hline  
         &   WD&FAD&  KL &  WD&FAD&KL\\ \hline 
         LZMidi&   \textbf{8.57}&$\textbf{0.69}$& \textbf{1.42} &  \textbf{8.39}& \textbf{0.64}&\textbf{1.37}\\ \hline 
 ASD3PM&  27.91&4.22&$2.29$ &  27.96&4.05&2.26\\\hline
    \end{tabular}
  
\end{table}

\subsection{Training, Generation Time, and Memory Usage}

\vspace{-10pt}

We record the training time, generation time (per sample), and memory usage of LZMidi to underscore its computational advantages over deep learning–based methods. Table~\ref{tab:train_gen_time} summarizes these metrics for different \(L\) settings. The metrics for our models are substantially lower than those of D3PM. For instance, while our training time is fixed at approximately one hour (reflecting our computational constraints), \cite{mittal2021symbolicmusicgenerationdiffusion} report a 6.5-hour training duration on an Nvidia Tesla V100 GPU.

\begin{table}[htbp]
    \centering
\caption{Training Time, Generation Time, Memory Usage}
\label{tab:train_gen_time}
    \begin{tabular}{|>{\centering\arraybackslash}p{0.15\linewidth}|>{\centering\arraybackslash}p{0.15\linewidth}|>{\centering\arraybackslash}p{0.15\linewidth}|>{\centering\arraybackslash}p{0.15\linewidth}|} \hline 
         &  Training Time (s)&  Generation Time (s/sample)& Model Size (MB)\\ \hline 
         LZMidi&  \textbf{107.7}&  \textbf{0.016}& \textbf{287.1}\\ \hline 
 ASD3PM& 3480& 5.4 &306.2\\\hline
    \end{tabular}  
\end{table}

\vspace{-10pt}

\subsection{Floating point operations (FLOPS) for ASD3PM Baseline}

\vspace{-10pt}

We analyze the computation (FLOPS) required for sequential MIDI generation using the ASD3PM baseline. We set the sequence length for the generated MIDI files to 256 timesteps as in the experiments. To evaluate the computational efficiency of our diffusion-based baseline model for symbolic music generation, we analyzed the floating-point operations per second (FLOPS) across different layers of the network during training and inference. Table~\ref{tab:flops} provides a breakdown of the total FLOPS and the corresponding CPU utilization metrics for key operations. 

\begin{table}[h!]
\centering
\caption{FLOPs and CPU Utilization of the Diffusion Baseline}
\label{tab:flops}
\resizebox{\columnwidth}{!}{%
\begin{tabular}{|l|c|c|c|c|}
\hline
\textbf{Operation} & \textbf{Calls} & \textbf{Total FLOPs} (MFLOPs) & \textbf{Self CPU \%} & \textbf{CPU Total Time (ms)} \\ \hline
\texttt{aten::addmm} & 145 & 620,622.774 & 37.36\% & 736.399 \\ \hline
\texttt{aten::bmm} & 48 & 12,884.902 & 6.91\% & 126.123 \\ \hline
\texttt{aten::mul} & 1 & 377.487 & 1.08\% & 21.089 \\ \hline
\texttt{aten::add} & 97 & 203.424 & 1.25\% & 14.255 \\ \hline
\texttt{DataParallel::forward} & 50 & 104.858 & 2.14\% & 18.332 \\ \hline
\texttt{aten::expand} & 243 & 0.00 & 0.56\% & 55.675 \\ \hline
\texttt{aten::reshape} & 245 & 0.00 & 0.88\% & 56.436 \\ \hline
\end{tabular}%
}
\end{table}


\vspace{-10pt}

\section{Conclusion}

\vspace{-10pt}

In this work, we introduced a novel approach to symbolic music generation using LZ78-based sequential probability assignment (SPA). Our method generates high-fidelity musical samples while substantially reducing computational requirements relative to state-of-the-art deep learning methods. Our experiments show that LZMidi effectively captures musical patterns, maintaining diversity and consistency metrics that rival or exceed those of diffusion-based models. Our experiments suggest that LZMidi rivals the diffusion baseline on all the aforementioned qualitative metrics (WD, FAD, KL) during both training and testing. Moreover, the method exhibits significant computational benefits—lower training/generation times and reduced memory usage—thus enabling efficient symbolic music generation on CPUs.

Future work will extend this study by training on longer sequences (e.g., 64-bar samples), as well as polyphonic sequences, and comparing results against fully trained diffusion models to assess scalability. 


\begin{remark}
Some of our generated samples (corresponding to the plots in \ref{fig:sample_all}) are attached \href{https://drive.google.com/drive/folders/1YglJn_KnBWZnze5xDQgHP5ufKqmZsaHn?usp=share_link}{here} for the reader's listening. 


\end{remark}

\bibliographystyle{plain}
\bibliography{references}  

\onecolumn

\appendix
\subsection{LZ78-Based Sequential Probability Assignment for Symbolic Music Generation}
\label{app:SPA}

In this work, we leverage the LZ78-based Sequential Probability Assignment (SPA) \cite{sagan2024familylz78baseduniversalsequential} for modeling symbolic music sequences. The model employs the incremental parsing algorithm of LZ78 to construct a prefix tree representation of an input training corpus of note events. Unlike conventional deep learning approaches that require extensive parameter optimization, the LZ78-SPA model learns symbol-by-symbol probability assignments incrementally and efficiently.

\subsubsection*{Preliminaries and Notation}

Let \(\{x_t\}_{t=1}^n\) be a sequence of discrete musical symbols drawn from a finite alphabet \(\mathcal{A}\), where \(|\mathcal{A}| = A < \infty.\) Each symbol \(x_t \in \mathcal{A}\) may represent a specific note or a rest indicator. The LZ78 parsing procedure incrementally partitions the sequence \(x^n = x_1 x_2 \cdots x_n\) into a set of phrases (nodes) that form a growing prefix tree.

We denote by \(Z(x^t)\) the set of phrases (including the empty root phrase) in the LZ78 parsing of the prefix \(x^t\). Each phrase is a node in the prefix tree. For a given symbol \(x_t\), let \(z(x_t)\) represent the phrase (node) associated with \(x_t\), and let \(z_c(x_{t-1})\) represent the LZ78 context of \(x_t\), i.e., the phrase prefix excluding the current symbol \(x_t\).

\subsubsection*{Universal Sequential Probability Assignment}

We define a sequential probability assignment (SPA) as a family of conditional distributions:
\begin{equation}
q = \{ q_t(\cdot \mid x^{t-1}) : t \geq 1 \}, \quad q_t(a | x^{t-1}) \in M(\mathcal{A}),
\end{equation}
where \(q_t(a | x^{t-1})\) is the probability of symbol \(a \in \mathcal{A}\) given the observed sequence \(x^{t-1}\).

Following \cite{sagan2024familylz78baseduniversalsequential}, our LZ78-SPA conditions the probability assignment for \(x_t\) on its LZ78 context \(z_c(x_{t-1})\). To introduce smoothness and avoid zero-probability assignments for unseen events, we employ a Dirichlet prior with parameter \(\gamma > 0\). This prior acts as an additive smoothing factor. Formally, define 
$N(a |x^{t-1}, z_c(x_{t-1}))$ as the count of symbol $a$ for phrases with prior context $(z_c(x_{t-1})$.

The LZ78-based SPA with a Dirichlet prior is given by:
\begin{equation}
    q^{LZ,\gamma}(a|x^{t-1})\triangleq \frac{N_{LZ}(a|x^{t-1})+\gamma}{\sum_{a'\in \mathcal{X}}N_{LZ}(a'|x^{t-1})+\gamma|\mathcal{X}|}.
    \label{eq:spa2}    
\end{equation}

Equivalently, since \(\sum_{b \in \mathcal{A}} N(b \mid x^{t-1}, z_c(x_{t-1}))\) is the number of times we have visited the context \(z_c(x_{t-1})\) in the prefix tree, this can be written as:
\[
q_t(a | x^{t-1}) = \frac{N(a | x^{t-1}, z_c(x_{t-1})) + \gamma}{N(\cdot | x^{t-1}, z_c(x_{t-1})) + \gamma A},
\]
where
\[
N(\cdot | x^{t-1}, z_c(x_{t-1})) = \sum_{b \in \mathcal{A}} N(b | x^{t-1}, z_c(x_{t-1})).
\]

\subsubsection*{Training Procedure}

Training the LZ78-SPA consists of two steps:

\begin{enumerate}
    \item \textbf{LZ78 Parsing:} Given the training corpus of symbolic music, we parse the entire training set using the LZ78 algorithm. This involves incrementally building a prefix tree, adding a new branch whenever a previously unseen context-symbol combination is encountered. The result is a tree structure \(Z(x^N)\) where \(N\) is the length of the entire training set.

    \item \textbf{Count Aggregation and Prior Incorporation:} For each node (phrase) in the LZ78 tree, we record the counts \(N(a \mid x^{t-1}, z_c(x_{t-1}))\) for all symbols \(a \in \mathcal{A}\) that follow the context \(z_c(x_{t-1})\). After parsing, these counts are fixed and used with the Dirichlet prior parameter \(\gamma\) to define the SPA.
\end{enumerate}

\subsubsection*{Loss Function (Log Loss)}

The quality of a sequential probability assignment is typically measured by the log loss. For a given test sequence \(x^n\), the log loss under the LZ78-SPA model is:
\[
L(x^n) = -\frac{1}{n} \sum_{t=1}^n \log q_t(x_t \mid x^{t-1}).
\]

This log loss corresponds to the negative log-likelihood per symbol and provides a measure of how well the learned model predicts the sequence. Lower values of \(L(x^n)\) indicate better predictive performance.

In practice, the LZ78-SPA often achieves asymptotically optimal log loss behavior when compared to Markovian or finite-state models \cite{sagan2024familylz78baseduniversalsequential}. For symbolic music generation, this translates into capturing the underlying repetitive and hierarchical patterns of musical structure while requiring significantly fewer computational resources than typical deep learning approaches.

\subsection{Proof of the Theorem}
\label{app:proof}
\begin{definition}
    Let $Q^m$ be the probability model induced by an LZ78 tree built with $m$ equal-length realizations $X^n \simiid P_{X^n}$ sampled from source $P$ over alphabet $\Acal$.
    Let $X^{(i),n}$ denote the $i$\textsuperscript{th} such sequence generated.
    No assumptions are placed on $P$.
\end{definition}
\begin{remark}
    In this setting, the depth of the LZ78 tree is upper-bounded by $n$.
\end{remark}
\begin{definition}[Symbol counts]
    $\Ccal(a|x^n)$ is the number of times that symbol $a$ appears in $x^n$.
\end{definition}
\begin{theorem}
    Assume that the SPA at each node of the LZ78 tree, $q$, satisfies $q(a|y^m) - \frac{\Ccal(a|y^m)}{m} \to 0$, for all individual sequences $\yv$.
    Then as $m\to\infty$,
    \[D(P_{X^n}\lVert Q^m_{X^n}) \convas 0.\]

    \begin{proof}
        First, we define some additional notation:
        \begin{itemize}
            \item When parsing the $m$\textsuperscript{th} sample at index $t$, denote the current node of the LZ78 tree by $z_t^m$.
            The subsequence of symbols seen at $z_t^m$ until time $t$ is denoted $\Scal(z_t^m, m, t)$.
            Denote the length of this subsequence by $\ell(z_t^m, m, t)$.
            \item The LZ78 SPA at sample $m$, step $t$ is denoted $q(\cdot|\Scal(z_t^m, m, t))$.
            The SPA for a node that has not yet see data is denoted $q(\cdot)$.
        \end{itemize} ·
        Consider any $Y^n \in \Acal^n$ such that $P_{X^n}(Y^n) > 0$.
        \[\log \frac{1}{Q_{X^n}^m(Y^n)} = \sum_{t=1}^n \log \frac{1}{q(Y_t|\Scal(z_t^m, m, t))}.\]

        \begin{fact}\label{fact:count-ratio-conv}
            Fix $t \geq 1$, $Y^{t-1} \in \Acal^n$ such that $P_{X^{t-1}}(Y^{t-1}) > 0$.
            Then, $\forall a \in \Acal$,
            \[\frac{\Ccal(a|\Scal(z_t^m, m, t)))}{\ell(z_t^m, m, t)} \convas P_{X_{t}|X^{t-1}}(a|Y^{t})\quad \text{as } m\to\infty.\]

            \begin{proof}
                Fix some $m > 0$, and define
                \[\Ccal_m(Y^{t-1}) = \sum_{i=1}^m \indic{X^{(i),t-1} = Y^{t-1}}.\]
                We can bound $\ell(z_t^m, m, t)$ by a constant plus $\Ccal_m(Y^{t-1})$ on both sides.
                Note that, for $\mathcal{C}_m(Y^{t-1}) \geq t$, no returns to the root occur before reaching $z_t^m$, so $z_t^m$ is a depth-$(t-1)$ corresponding directly to the prefix $Y^{t-1}$.
                Otherwise, $z_t^m$ is some node encountered after a return to the root, in which case we cannot say much about $\ell(z_t^m, m, t)$ relative to $\mathcal{C}_m(Y^{t-1})$.

                By the law of large numbers, $\frac{1}{m} \Ccal_m(Y^{t-1}) \convas P_{X^{t-1}}(Y^{t-1}) > 0$ as $m \to \infty$.
                Therefore, there almost surely exists some $M$ such that $\Ccal_M(Y^{t-1}) \geq t$.
                From this point, consider $m > M$.
                
                For a lower bound on $\ell(z_t^m, m, t)$, the node $z_t^m$ was visited for all but maybe the first $t$ times $Y^t$ was seen.
                For an upper bound, we consider all the possible times $z_t^m$ was visited that do not correspond to $Y^{t-1}$, \ie, times  $z_t^m$ was visited after a return to the root.
                There is exactly one return to the root for every leaf of the tree, so the number of extra visits is bounded.
                So, $\forall m > M$, almost surely $\ell(z_t^m, m, t) = \Ccal_m(Y^{t-1}) + O(1)$.

                By the same logic, $\Ccal(a|\Scal(z_t^m), m, t)) = \Ccal_m(Y^{t-1} \,^\frown a) + O(1)$, where $\,^\frown$ represents sequence concatenation.

                By the law of large numbers, $\frac{1}{m} \Ccal_m(Y^{t-1}) = P_{X^{t-1}}(Y^{t-1}) + o_p(1)$, and analogously for $\frac{1}{m} \Ccal_m(Y^{t-1} \,^\frown a)$.
                As a result,
                \[\frac{\Ccal(a|\Scal(z_t^m, m, t)))}{\ell(z_t^m, m, t)} = \frac{\frac{1}{m}\left(\Ccal_m(Y^{t-1} \,^\frown a) + O(1) \right)}{\frac{1}{m}\left(\Ccal_m(Y^{t-1}) + O(1)\right)} = \frac{P_{X^{t}}(Y^{t-1} \,^\frown a) + o_p(1)}{P_{X^{t-1}}(Y^{t-1}) + o_p(1)} \convas P_{X_{t}|X^{t-1}}(a|Y^{t-1}),\]
                by Slutsky's theorem.
            \end{proof}
        \end{fact}
        \begin{corollary}
            We know that, almost surely, for sufficiently large $m$ $\ell(z_t^m, m, t) = \Ccal_m(Y^{t-1}) + O(1)$.
            So, by the law of large numbers, $\ell(z_t^m, m, t)$ almost surely grows unbounded, for any $Y^{t-1}$ with non-zero measure under $P$.
        \end{corollary}
        \begin{corollary}\label{cor:node-spa-convergence}
            By assumption, $\forall\, Y^{t-1} \in \Acal^{t-1}$ with nonzero measure under $P$, 
            \[q(a|\Scal(z_t^m, m, t)) \to \frac{\Ccal(a|\Scal(z_t^m, m, t)))}{\ell(z_t^m, m, t)}, \quad\text{as}\quad \ell(z_t^m, m, t) \to \infty.\]
            Applying the fact $\ell(z_t^m, m, t)\convas \infty$, along with \prettyref{fact:count-ratio-conv}, as $m\to\infty$,
            \[q(a|\Scal(z_t^m, m, t)) \convas P_{X_{t}|X^{t-1}}(a|Y^{t-1}).\]
        \end{corollary}
        By \prettyref{cor:node-spa-convergence}, the continuous mapping theorem, and Slutsky's theorem (as $n < \infty$ is a fixed quantity), for any fixed $Y^n$ with nonzero measure under $P$,
        \[\log \frac{1}{Q_{X^n}^m(Y^n)} = \sum_{t=1}^n \log \frac{1}{q(Y_t|\Scal(z_t^m, m, t))} \convas \sum_{t=1}^n \log \frac{1}{P_{X_{t}|X^{t-1}}(Y_{t}|Y^{t-1})} = \log \frac{1}{P_{X^n}(Y^n)}.\]
        Applying this to the relative entropy,
        \[D(P_{X^n}\lVert Q^m_{X^n}) = \Ebb \left[ \log P_{X^n}(X^n)\right] + \Ebb \left[ \log Q^m_{X^n}(X^n)\right] = \Ebb \left[ \log P_{X^n}(X^n)\right] + \sum_{Y^n : P_{X^n}(Y^n) > 0} P_{X^n}(Y^n) \log \frac{1}{Q^m_{X^n}(Y^n)}. \]
        Applying Slutsky's theorem (using the fact that the summation has a fixed, finite number of terms),
        \[\sum_{Y^n : P_{X^n}(Y^n) > 0} P_{X^n}(Y^n) \log \frac{1}{Q^m_{X^n}(Y^n)} \convas \Ebb \left[ \log P_{X^n}(X^n)\right],\]
        and, therefore,
        \[D(P_{X^n}\lVert Q^m_{X^n}) \convas \Ebb \left[ \log P_{X^n}(X^n)\right] - \Ebb \left[ \log P_{X^n}(X^n)\right] = 0.\]
    \end{proof}
\end{theorem}

Additionally, in our future work, we intend to compare LZMidi to other symbolic music baselines (e.g., n-gram or transformer models) for a more direct comparison in unconditional settings.

\subsection{Fully-trained ASD3PM results}
\label{sec:fullytrained}

In addition to the partially-trained diffusion baseline (ASD3PM) results presented in Section \ref{sec:results} (Plasser et al. \cite{plasser2023discretediffusionprobabilisticmodels}), we provide here the results for the fully-trained ASD3PM diffusion model. We additionally emphasized several key experimental differences between their setup and ours in 1) Sequence Length (256 vs. 1024 tokens) and 2) Hardware \& Training time.

Due to these differences, the following results are \textbf{not} directly comparable but instead provide additional context regarding the performance achieved by a longer-trained diffusion approach with extended sequences:

\begin{table}[h!]
\centering
\begin{tabular}{lcccccccc}
\hline
Setting & \multicolumn{4}{c}{Unconditional} \\ \hline
& \multicolumn{2}{c}{Pitch} & \multicolumn{2}{c}{Duration}\\ \hline
\textbf{Metric} & C & Var & C & Var \\ \hline
\textbf{ASD3PM} & \textbf{0.992} & \textbf{0.920} & \textbf{0.993} & \textbf{0.937} \\ \hline
\end{tabular}
\caption{Fully-trained metrics for Melody 64 bar setting.}
\label{tab:schmubert-melody64}
\end{table}

These fully-trained results serve to contextualize the partially-trained ASD3PM baseline discussed in the main text.

\end{document}